\documentclass[twoside]{dis09}
\usepackage[latin1]{inputenc}
\usepackage[dvips]{graphicx,epsfig,color}
\usepackage{wrapfig,rotating}
\usepackage{amssymb,amsmath,array}

\newcommand\as{\alpha_{\mathrm{S}}}
\def\ptmin{p_{T{\rm min}}}
\def\ptmax{p_{T{\rm max}}}
\begin{document}
\title{The Drell-Yan process in NNLO QCD\footnote{Invited talk given at the XVII International Workshop on Deep-Inelastic Scattering and Related Subjects DIS 2009, 26-30 April 2009, Madrid}}

\author{Massimiliano Grazzini
%
%
\vspace{.3cm}\\
%
INFN, Sezione di Firenze and
Dipartimento di Fisica, Universit\`a di Firenze,\\
I-50019 Sesto Fiorentino, Florence, Italy\\
%
}

\maketitle

\begin{abstract}
We consider the production of $W$ and $Z$ bosons
in hadron collisions. We present a selection of numerical results obtained through a fully exclusive calculation up to next-to-next-to-leading
order (NNLO) in QCD perturbation theory.
We include 
the $\gamma$--$Z$ interference, finite-width effects, the leptonic decay of the
vector bosons and the corresponding spin correlations.
The calculation is completely realistic, since it allows us
to apply arbitrary kinematical cuts
on the final-state leptons and the associated partons, and 
to compute the corresponding
distributions in the form of bin histograms.

\end{abstract}

The production of lepton pairs 
through the Drell--Yan (DY) mechanism \cite{Drell:1970wh}
was the first process where parton model ideas developed for deep inelastic scattering
were applied to hadronic collisions, and lead to the discovery of $W$ and $Z$ bosons \cite{Arnison:1983rp,Banner:1983jy}.

It is thus not surprizing that
the production of vector bosons
is central in physics studies at hadron colliders.
These processes
have large production rates and clean experimental signatures, given the presence
of at least one high-$p_T$ lepton in the final state.
Studies of the production of $W$ bosons at the Tevatron lead to precise determinations
of the $W$ mass and width \cite{:2008ut}.
Vector boson production
is also expected to provide standard candles for detector calibration
during the first stage of the LHC running. 

For these reasons it is important to have accurate
theoretical predictions for the vector-boson production cross sections
and the associated distributions,
and such a task requires
detailed computations of radiative corrections.
The QCD corrections to the total cross section \cite{Hamberg:1990np}
and to the rapidity distribution \cite{Anastasiou:2003yy,Anastasiou:2003ds} of the vector boson 
are known up to the next-to-next-to-leading order (NNLO) in the strong coupling
$\as$. The fully exclusive NNLO calculation, including the leptonic decay 
of the vector boson, has been completed more recently \cite{Melnikov:2006di,Catani:2009sm}.
Full electroweak corrections at ${\cal O}(\alpha)$
have been computed
for both $W$ \cite{ewW} and $Z$ production~\cite{ewZ}.

In this contribution we discuss a recent computation of the NNLO QCD corrections
to vector boson production in hadron collisions \cite{Catani:2009sm}.

The evaluation of higher-order QCD corrections to hard-scattering processes
is well known to be a hard task.
The presence of infrared (IR) singularities at intermediate stages of the calculation prevents
a straightforward implementation of numerical techniques.
In particular, NNLO {\em differential} calculations are a rarity due to their 
substantial technical complications.
In $e^+e^-$ collisions, NNLO differential cross sections are known only for 
two~\cite{Anastasiou:2004qd,Weinzierl:2006ij} and three jet production~\cite{threejets,Weinzierl:2008iv}.
At hadron colliders fully differential cross-sections have been computed only 
for Higgs production in gluon fusion~\cite{Anastasiou:2004xq,Catani:2007vq,Anastasiou:2007mz,Grazzini:2008tf}, 
and the Drell-Yan process~\cite{Melnikov:2006di,Catani:2009sm}. 
It is interesting to note that the amplitudes relevant for vector boson production
at NNLO have been known for at least 15 years \cite{Hamberg:1990np}
before the first fully exclusive computation could be completed \cite{Melnikov:2006di}.

\begin{wrapfigure}{r}{0.5\columnwidth}
\centerline{\includegraphics[width=0.45\columnwidth]{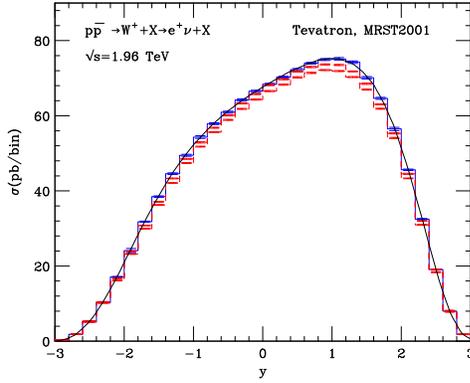}}
\caption{Rapidity distribution of the $W^+$ boson at the Tevatron. The NNLO result (blue) is compared to the NLO band (red) and to the NNLO prediction of Ref.~\cite{Anastasiou:2003ds}.}\label{fig:y34W}
\end{wrapfigure}
The calculation \cite{Catani:2009sm} we discuss here is based on
an extension of the subtraction formalism \cite{Frixione:1995ms,Catani:1996vz}
to NNLO that can be applied to the production of colourless high-mass system in hadron collisions \cite{Catani:2007vq}.
The calculation parallels the one recently completed
for Higgs boson production \cite{Catani:2007vq,Grazzini:2008tf},
and it is performed by using the same method.
We include
the $\gamma$--$Z$ interference, finite-width effects, the leptonic decay of the
vector bosons and the corresponding spin correlations.

In the following
we present some
numerical results for $W$ and $Z$ production at Tevatron energies.
We consider $n_F=5$ massless quarks in the initial state,
and, in the case of
$W^\pm$ production,
we use the (unitarity constrained) CKM matrix elements $V_{ud}=0.97419$, $V_{us}=0.2257$, $V_{ub}=0.00359$,
$V_{cd}=0.2256$, $V_{cs}=0.97334$, $V_{cb}=0.0415$ from the PDG 2008 \cite{Amsler:2008zzb}. 
In the case of $Z$ production, 
additional Feynman diagrams with fermionic triangles 
should be taken into account.
Their contribution cancels out for
each isospin multiplet when massless quarks are considered. The effect of
a finite top-quark mass
in the third generation has been considered and found extremely small \cite{Dicus:1985wx},
so it 
is 
neglected here.

As for the electroweak couplings, we use the so called $G_\mu$ scheme,
where the input parameters are $G_F$ , $m_Z$, $m_W$. In particular we 
use the values
$G_F = 1.16637\times 10^{-5}$~GeV$^{-2}$,
$m_Z = 91.1876$~GeV, $\Gamma_Z=2.4952$~GeV, $m_W = 80.398$~GeV
and $\Gamma_W=2.141$~GeV.
We use the 
MSTW2008 \cite{Martin:2009iq} sets
of parton distributions, with
densities and $\as$ evaluated at each corresponding order
(i.e., we use $(n+1)$-loop $\as$ at N$^n$LO, with $n=0,1,2$). The 
renormalization and factorization scales are fixed to the value 
$\mu_R=\mu_F=m_V$, where $m_V$ is the mass of the vector boson.

\begin{wrapfigure}{r}{0.5\columnwidth}
\centerline{\includegraphics[width=0.45\columnwidth]{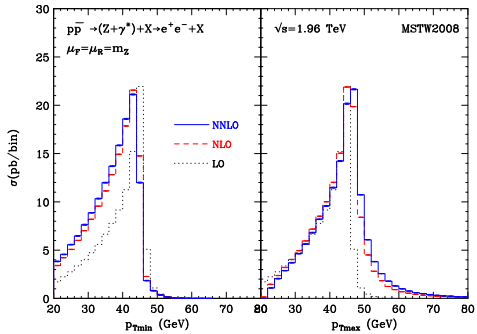}}
\caption{Distributions in $\ptmin$ and $\ptmax$ for the $Z$ signal at 
the Tevatron.}\label{fig:ptminmax}
\end{wrapfigure}
We start the presentation of our results by considering
the production of an on-shell $W^+$ boson
at the Tevatron.
When no cuts are applied our numerical program allows an independent computation of the rapidity distribution
of a vector boson up to NNLO \cite{Anastasiou:2003ds}.
To compare with Ref.~\cite{Anastasiou:2003ds},
in Fig.~\ref{fig:y34W} we show the rapidity distribution
of the $W^+$ obtained by using the MRST2001 partons \cite{Martin:2001es,Martin:2002dr}.
The blue histogram is the NNLO prediction; in red we also show the NLO band, obtained by varying $\mu_F=\mu_R$ between $1/2 m_W$ and $2m_W$.
The solid curve is the (scaled) NNLO prediction extracted from Fig.~10 of Ref.~\cite{Anastasiou:2003ds}.
The two NNLO results appear to be in good agreement.
\begin{wrapfigure}{r}{0.5\columnwidth}
\centerline{\includegraphics[width=0.45\columnwidth]{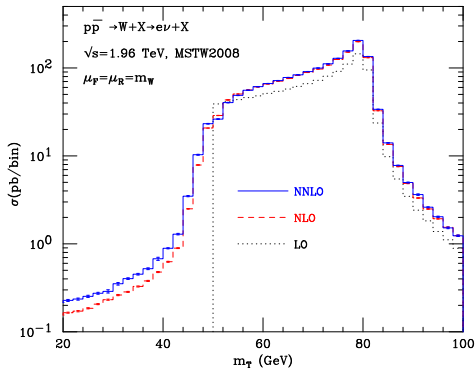}}
\caption{Distributions in $\ptmin$ and $\ptmax$ for the $Z$ signal at 
the Tevatron.}\label{fig:tmass}
\end{wrapfigure}
We next consider the production of $e^+e^-$ pairs from $Z/\gamma^*$ bosons at the Tevatron.
For each event, we classify the lepton transverse momenta according to their
minimum and maximum values,  
$\ptmin$ and $\ptmax$.
The 
leptons
are required to have a minimum $p_T$ of 20~GeV and pseudorapidity $|\eta|<2$.
Their invariant mass is required to be in the range 70~GeV~$<m_{\,e^+e^-}<110$~GeV.
The accepted cross sections are $\sigma_{LO}=103.37 \pm 0.04$~pb,
$\sigma_{NLO}=140.43 \pm 0.07$~pb and $\sigma_{NNLO}=143.86 \pm 0.12$~pb.

In Fig.~\ref{fig:ptminmax} we plot
the distributions in $\ptmin$ and $\ptmax$ at LO, NLO and NNLO.
We note that at LO the $\ptmin$ and $\ptmax$ distributions are kinematically
bounded by $p_T\le Q_{\rm max}/2$, where $Q_{\rm max}=110$~GeV is the maximum allowed invariant mass of the $e^+e^-$ pairs.
The NNLO corrections have a visible impact on the shape of the $\ptmin$ and $\ptmax$ distribution and
make the $\ptmin$ distribution softer, and the $\ptmax$ distribution harder.

We finally consider the production of a charged lepton plus missing $p_T$ through the decay of a $W$ boson ($W=W^+,W^-$) at the Tevatron.
The charged lepton is required to have $p_T>20$~GeV and $|\eta|<2$ and the missing $p_T$
of the event should be larger than 25~GeV. We define the transverse mass of the
event as $m_T=\sqrt{2p_T^lp_T^{\rm miss}(1-\cos\phi)}$, where $\phi$ is the angle between the
the $p_T$ of the lepton and the missing $p_T$. The accepted cross sections are
$\sigma_{LO}=1.161 \pm 0.001$~nb,
$\sigma_{NLO}=1.550\pm 0.001$~nb and $\sigma_{NNLO}=1.586 \pm 0.002$~nb.
In Fig.~\ref{fig:tmass} we show the $m_T$ distribution at LO, NLO and NNLO.
We note that at LO the distribution has a kinematical boundary at $m_T=50$~GeV.
This is due to the fact that at LO the $W$ is produced with zero 
transverse momentum:
therefore,
the requirement $p_T^{\rm miss}>25$~GeV sets
$m_T\geq 50$~GeV. 
Around the region where $m_T=50$~GeV there are perturbative 
instabilities in going from LO to NLO and to NNLO.
The origin of these perturbative instabilities is 
well 
known 
\cite{Catani:1997xc}:
since the LO spectrum
is kinematically bounded by $m_T\geq 50$~GeV,
each higher-order perturbative contribution produces
(integrable) logarithmic singularities in the vicinity of
the boundary.
We also note that, below the boundary, the NNLO corrections to the NLO result
are large; for example, they are about 
$+40$\% at $m_T\sim 30$~GeV. This is not unexpected, 
since in this region of transverse masses, the ${\cal O}(\as)$ result corresponds
to the calculation at 
the first perturbative order and, therefore, our ${\cal O}(\as^2)$ result 
is actually only a calculation at the NLO level of perturbative accuracy.

We have discussed some selected results 
of a calculation of $W$ and $Z$ boson production
up to NNLO in QCD perturbation theory. 
The calculation is
implemented in a parton level event generator and
it is particularly suitable for practical applications
to the computation of distributions in the form of bin histograms.
A public version of our program will be made available in the near future.

\begin{footnotesize}

\end{footnotesize}


\begin{thebibliography}{99}
\bibitem{Drell:1970wh}
  S.~D.~Drell and T.~M.~Yan,
  Phys.\ Rev.\ Lett.\  {\bf 25} (1970) 316
  [Erratum-ibid.\  {\bf 25} (1970) 902].

\bibitem{Arnison:1983rp}
  G.~Arnison {\it et al.}  [UA1 Collaboration],
  Phys.\ Lett.\  B {\bf 122} (1983) 103,
Phys.\ Lett.\  B {\bf 126} (1983) 398.

\bibitem{Banner:1983jy}
  M.~Banner {\it et al.}  [UA2 Collaboration],
  Phys.\ Lett.\  B {\bf 122} (1983) 476;
  P.~Bagnaia {\it et al.}  [UA2 Collaboration],
Phys.\ Lett.\  B {\bf 129} (1983) 130.

\bibitem{:2008ut}
See e.g. The Tevatron Electroweak Working Group for the CDF and D0 Collaborations,
{\em Combination of CDF and D0 results on the $W$ boson mass and width}, preprint FERMILAB-TM-2415, arXiv:0808.0147 [hep-ex].


\bibitem{Hamberg:1990np}
 R.~Hamberg, W.~L.~van Neerven and T.~Matsuura,
 Nucl.\ Phys.\  B {\bf 359} (1991) 343
 [Erratum-ibid.\  B {\bf 644} (2002) 403];
 R.~V.~Harlander and W.~B.~Kilgore,
 Phys.\ Rev.\ Lett.\  {\bf 88} (2002) 201801.

\bibitem{Anastasiou:2003yy}
  C.~Anastasiou, L.~J.~Dixon, K.~Melnikov and F.~Petriello,
  Phys.\ Rev.\ Lett.\  {\bf 91} (2003) 182002,
[arXiv:hep-ph/0306192].



\bibitem{Anastasiou:2003ds}
C.~Anastasiou, L.~J.~Dixon, K.~Melnikov and F.~Petriello,
Phys.\ Rev.\ D {\bf 69} (2004) 094008,
[arXiv:hep-ph/0312266].


\bibitem{Melnikov:2006di}
  K.~Melnikov and F.~Petriello,
  Phys.\ Rev.\ Lett.\  {\bf 96} (2006) 231803,
Phys.\ Rev.\ D {\bf 74} (2006) 114017.

\bibitem{Catani:2009sm}
  S.~Catani, L.~Cieri, G.~Ferrera, D.~de Florian and M.~Grazzini,
  arXiv:0903.2120 [hep-ph].



\bibitem{ewW}
S.~Dittmaier and M.~Kramer,
Phys.\ Rev.\  D {\bf 65} (2002) 073007;
U.~Baur and D.~Wackeroth,
Phys.\ Rev.\  D {\bf 70} (2004) 073015;
V.~A.~Zykunov,
Phys.\ Atom.\ Nucl.\  {\bf 69} (2006) 1522
[Yad.\ Fiz.\  {\bf 69} (2006) 1557];
A.~Arbuzov et al.,
Eur.\ Phys.\ J.\  C {\bf 46} (2006) 407
[Erratum-ibid.\  C {\bf 50} (2007) 505];
C.~M.~Carloni Calame, G.~Montagna, O.~Nicrosini and A.~Vicini,
JHEP {\bf 0612} (2006) 016.



\bibitem{ewZ}
U.~Baur, O.~Brein, W.~Hollik, C.~Schappacher and D.~Wackeroth,
Phys.\ Rev.\  D {\bf 65} (2002) 033007;
V.~A.~Zykunov,
Phys.\ Rev.\  D {\bf 75} (2007) 073019;
C.~M.~Carloni Calame, G.~Montagna, O.~Nicrosini and A.~Vicini,
JHEP {\bf 0710} (2007) 109;
A.~Arbuzov et al.,
Eur.\ Phys.\ J.\  C {\bf 54} (2008) 451.




\bibitem{Anastasiou:2004qd}
  C.~Anastasiou, K.~Melnikov and F.~Petriello,
  Phys.\ Rev.\ Lett.\  {\bf 93} (2004) 032002.



\bibitem{Weinzierl:2006ij}
  S.~Weinzierl,
  Phys.\ Rev.\  D {\bf 74} (2006) 014020
  [arXiv:hep-ph/0606008].


\bibitem{threejets}
  A.~Gehrmann-De Ridder, T.~Gehrmann, E.~W.~N.~Glover and G.~Heinrich,
  Phys.\ Rev.\ Lett.\  {\bf 99} (2007) 132002,
  JHEP {\bf 0711} (2007) 058,
JHEP {\bf 0712} (2007) 094,
Phys.\ Rev.\ Lett.\  {\bf 100} (2008) 172001;

\bibitem{Weinzierl:2008iv}
  S.~Weinzierl,
  Phys.\ Rev.\ Lett.\  {\bf 101} (2008) 162001.







\bibitem{Anastasiou:2004xq}
  C.~Anastasiou, K.~Melnikov and F.~Petriello,
  Phys.\ Rev.\ Lett.\  {\bf 93} (2004) 262002,
Nucl.\ Phys.\ B {\bf 724} (2005) 197.

\bibitem{Catani:2007vq}
  S.~Catani and M.~Grazzini,
  Phys.\ Rev.\ Lett.\  {\bf 98} (2007) 222002.


\bibitem{Anastasiou:2007mz}
  C.~Anastasiou, G.~Dissertori and F.~Stockli,
  JHEP {\bf 0709} (2007) 018.


\bibitem{Grazzini:2008tf}
  M.~Grazzini,
  JHEP {\bf 0802} (2008) 043.




\bibitem{Frixione:1995ms}
S.~Frixione, Z.~Kunszt and A.~Signer,
Nucl.\ Phys.\ B {\bf 467} (1996) 399;
S.~Frixione,
Nucl.\ Phys.\ B {\bf 507} (1997) 295.

\bibitem{Catani:1996vz}
S.~Catani and M.~H.~Seymour,
Nucl.\ Phys.\ B {\bf 485} (1997) 291
[Erratum-ibid.\ B {\bf 510} (1997) 503].


\bibitem{Amsler:2008zzb}
  C.~Amsler {\it et al.}  [Particle Data Group],
  Phys.\ Lett.\  B {\bf 667} (2008) 1.



\bibitem{Martin:2009iq}
  A.~D.~Martin, W.~J.~Stirling, R.~S.~Thorne and G.~Watt,
  report IPPP/08/190 [arXiv:0901.0002].




\bibitem{Martin:2001es}
  A.~D.~Martin, R.~G.~Roberts, W.~J.~Stirling and R.~S.~Thorne,
  Eur.\ Phys.\ J.\  C {\bf 23} (2002) 73
  [arXiv:hep-ph/0110215].

\bibitem{Martin:2002dr}
  A.~D.~Martin, R.~G.~Roberts, W.~J.~Stirling and R.~S.~Thorne,
  Phys.\ Lett.\  B {\bf 531} (2002) 216
  [arXiv:hep-ph/0201127].






\bibitem{Dicus:1985wx}
  D.~A.~Dicus and S.~S.~D.~Willenbrock,
  Phys.\ Rev.\  D {\bf 34} (1986) 148.




\bibitem{Catani:1997xc}
  S.~Catani and B.~R.~Webber,
  JHEP {\bf 9710} (1997) 005.

\end{thebibliography}
\end{document}